\documentclass[jog]{igs}
\usepackage[utf8]{inputenc}

\usepackage{graphicx}
\usepackage{siunitx}
\usepackage{subcaption}

\usepackage{hyperref}
\usepackage{doi}

\begin{document}

\title[~]{Radiofrequency Ice Dielectric Measurements at Summit Station, Greenland}


\author[Aguilar and others]{
  J.~A.~Aguilar$^{1}$,
  P.~Allison$^{2}$,
  D.~Besson$^{3}$,
  A.~Bishop$^{4}$,
  O.~Botner$^{5}$,
  S.~Bouma$^{6}$,
  S.~Buitink$^{7}$,
  M.~Cataldo$^{6}$,
  B.~A.~Clark$^{8}$,
  K.~Couberly$^{3}$,
  Z.~Curtis-Ginsberg$^{9}$,
  P.~Dasgupta$^{1}$,
  S.~de~Kockere$^{10}$,
  K.~D.~de~Vries$^{10}$,
  C.~Deaconu$^{9}$,
  M.~A.~DuVernois$^{4}$,
  A.~Eimer$^{6}$,
  C.~Glaser$^{5}$,
  A.~Hallgren$^{5}$,
  S.~Hallmann$^{11}$,
  J.~C.~Hanson$^{12}$,
  B.~Hendricks$^{13}$,
  J.~Henrichs$^{11,6}$,
  N.~Heyer$^{5}$,
  C.~Hornhuber$^{3}$,
  K.~Hughes$^{9}$,
  T.~Karg$^{11}$,
  A.~Karle$^{4}$,
  J.~L.~Kelley$^{4}$,
  M.~Korntheuer$^{1}$,
  M.~Kowalski$^{11,14}$,
  I.~Kravchenko$^{15}$,
  R.~Krebs$^{13}$,
  R.~Lahmann$^{6}$,
  U.~Latif$^{10}$,
  J.~Mammo$^{15}$,
  M.~J.~Marsee$^{16}$,
  Z.~S.~Meyers$^{11,6}$,
  K.~Michaels$^{9}$,
  K.~Mulrey$^{17}$,
  M.~Muzio$^{13}$,
  A.~Nelles$^{11,6}$,
  A.~Novikov$^{18}$,
  A.~Nozdrina$^{3}$,
  E.~Oberla$^{9}$,
  B.~Oeyen$^{19}$,
  I.~Plaisier$^{6,11}$,
  N.~Punsuebsay$^{18}$,
  L.~Pyras$^{11,6}$,
  D.~Ryckbosch$^{19}$,
  O.~Scholten$^{10,20}$,
  D.~Seckel$^{18}$,
  M.~F.~H.~Seikh$^{3}$,
  D.~Smith$^{9}$,
  J.~Stoffels$^{10}$,
  D.~Southall$^{9}$,
  K.~Terveer$^{6}$,
  S.~Toscano$^{1}$,
  D.~Tosi$^{4}$,
  D.~J.~Van~Den~Broeck$^{10,7}$,
  N.~van~Eijndhoven$^{10}$,
  A.~G.~Vieregg$^{9}$,
  J.~Z.~Vischer$^{6}$,
  C.~Welling$^{9}$,
  D.~R.~Williams$^{16}$,
  S.~Wissel$^{13}$,
  R.~Young$^{3}$,
  A.~Zink$^{6}$
}
\affiliation{
 $^{1}$Universit\'e Libre de Bruxelles, Science Faculty CP230, B-1050 Brussels, Belgium,\\
 $^{2}$Dept.\ of Physics, Center for Cosmology and AstroParticle Physics, Ohio State University, Columbus, OH 43210, USA,\\
 $^{3}$University of Kansas, Dept.\ of Physics and Astronomy, Lawrence, KS 66045, USA,\\
 $^{4}$Wisconsin IceCube Particle Astrophysics Center (WIPAC) and Dept.\ of Physics, University of Wisconsin-Madison, Madison, WI 53703,  USA,\\
 $^{5}$Uppsala University, Dept.\ of Physics and Astronomy, Uppsala, SE-752 37, Sweden,\\
 $^{6}$Erlangen Center for Astroparticle Physics (ECAP), Friedrich-Alexander-University Erlangen-N\"urnberg, 91058 Erlangen, Germany,\\
 $^{7}$Vrije Universiteit Brussel, Astrophysical Institute, Pleinlaan 2, 1050 Brussels, Belgium,\\
 $^{8}$Dept.\ of Physics and Astronomy, Michigan State University, East Lansing MI 48824, USA,\\
 $^{9}$Dept.\ of Physics, Enrico Fermi Inst., Kavli Inst.\ for Cosmological Physics, University of Chicago, Chicago, IL 60637, USA,\\
 $^{10}$Vrije Universiteit Brussel, Dienst ELEM, B-1050 Brussels, Belgium,\\
 $^{11}$Deutsches Elektronen-Synchrotron DESY, Platanenallee 6, 15738 Zeuthen, Germany,\\
 $^{12}$Whittier College, Whittier, CA 90602, USA,\\
 $^{13}$Dept.\ of Physics, Dept.\ of Astronomy \& Astrophysics, Penn State University, University Park, PA 16801, USA,\\
 $^{14}$Institut für Physik, Humboldt-Universit\"at zu Berlin, 12489 Berlin, Germany,\\
 $^{15}$Dept.\ of Physics and Astronomy, Univ.\ of Nebraska-Lincoln, NE, 68588, USA,\\
 $^{16}$Dept.\ of Physics and Astronomy, University of Alabama, Tuscaloosa, AL 35487, USA,\\
 $^{17}$Dept.\ of Astrophysics/IMAPP, Radboud University, PO Box 9010, 6500 GL, The Netherlands,\\
 $^{18}$Dept.\ of Physics and Astronomy, University of Delaware, Newark, DE 19716, USA,\\
 $^{19}$Ghent University, Dept.\ of Physics and Astronomy, B-9000 Gent, Belgium,\\
 $^{20}$Kapteyn Institute, University of Groningen, Groningen, The Netherlands
}

\begin{frontmatter}

\maketitle
\begin{abstract}

We recently reported \citep{aguilar2022situ} on the radio-frequency attenuation length of cold polar ice at Summit Station, Greenland, based on bistatic radar measurements of radio-frequency bedrock echo strengths taken during the summer of 2021. Those data also include echoes attributed to stratified impurities or dielectric discontinuities within the ice sheet (``layers"), which allow studies of a) estimation of the relative contribution of coherent (discrete layers, e.g.) vs. incoherent (bulk volumetric, e.g.) scattering, b) the magnitude of internal layer reflection coefficients, c) limits on the azimuthal asymmetry of reflections (`birefringence'), and d) limits on signal dispersion in-ice over a bandwidth of $\sim$100 MHz. We find that i) after averaging 10000 echo triggers, reflected signal observable over the thermal floor (to depths of approximately 1500 m) are consistent with being entirely coherent, ii) internal layer reflection coefficients are measured at approximately --60 to --70 dB, iii) birefringent effects for vertically propagating signals are smaller by an order of magnitude relative to comparable studies performed at South Pole, and iv) within our experimental limits, glacial ice is non-dispersive over the frequency band relevant for neutrino detection experiments.
\message{, and c) a precise measurement of the asymptotic refractive index of the Greenland ice sheet, by correlating our observed radar echoes with the known depth of impurity layers catalogued for the GISP core.} 
\end{abstract}
\end{frontmatter}

\section{Introduction}
The emerging field of `multi-messenger astronomy' has, as one of its primary goals, elucidation of the detailed mechanisms that drive the most energetic accelerators in the Universe\citep{meszaros2019multi}. Synthesis of measurements from contemporary gamma-ray, charged cosmic-ray, neutrino, and even gravitational wave observatories allows considerably enhanced refinement of astrophysical models than afforded by one cosmic messenger alone. The Earth's polar ice sheets comprise a large and nearly homogeneous target volume for cosmic neutrinos, which can then be detected via the electromagnetic Cherenkov radiation following an interaction of a neutrino with an ice molecule target. At the highest neutrino energies, detection of coherent radio-frequency emissions is perhaps the most cost-effective approach given the high transparency of cold polar ice to radio waves \citep{barwick_besson_gorham_saltzberg_2005,BESSON2008130,ALLISON2012457,Avva:2014ena,barrella_barwick_saltzberg_2011,hanson_2015,aguilar2022situ}. Site-specific ice dielectric measurements are required to quantify the expected neutrino detection rate at a given location. The recently-initiated Radio Neutrino Observatory in Greenland (RNO-G), located at Summit Station in Greenland, seeks first-ever detection of so-called `cosmogenic' neutrinos \citep{aguilar2021design}. In tandem with the deployment of the first radio receivers in the summer of 2021, an extensive data sample was accumulated to quantify the ice radio-glaciological characteristics at Summit Station. Below, we detail measurements, not only of the bulk ice attenuation length, but also of the characteristics of radio scattering layers embedded within the ice sheet itself, including the polarization dependence of the scattering (`birefringence').

\section{Measurements}
Our 2021 ice calibration campaign had four primary measurement goals, seeking to quantify the following:
\begin{enumerate}
\item Numerically, the attenuation length ($L_\alpha$) determines the neutrino detector effective volume 
for in-ice radio receiver arrays at high neutrino energies ($E_\nu\gtrsim $1~EeV). 
\item Since a variable refractive index profile results in curved, rather than rectilinear ray trajectories, some neutrino interaction vertices are inaccessible to shallow radio receivers. The refractive index profile with depth therefore determines the degree of `shadowing' for signals arriving from primarily horizontal directions; this effect is most important at the lowest detectable energy neutrinos using the radio technique ($E_\nu \lesssim$100 PeV). 
\item Anisotropy in signal velocity with electric-field $\vec{E}$ polarization direction (``birefringence'') leads to signal splitting and rotation of polarization vectors, depending on propagation and polarization directions, with signal arrival time staggers of order 1--10 ns per km of distance propagated. 
\item Reflection of radio-frequency signals by internal layers can both complicate signal recognition and also present a potential background channel when above-surface radio-frequency noise propagating into the ice may be reflected upwards towards a shallower receiver (Rx) antenna. 
\end{enumerate}

\subsection{Summit Station Radioglaciology}
\begin{figure}
\includegraphics[width=\columnwidth]{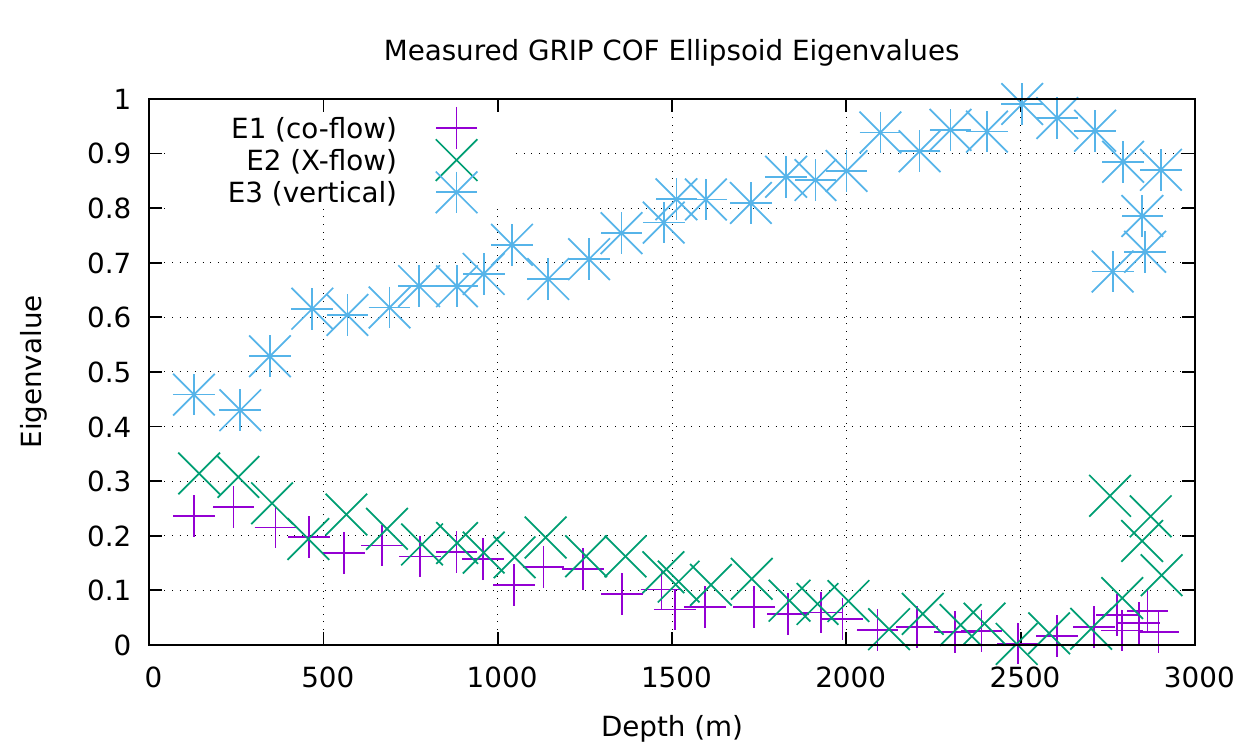}
\caption{Published magnitudes of horizontal (E1 and E2) vs.\ vertical (E3) components of crystal orientation fabric tensor for data \citep{thorsteinsson1997textures}, referenced relative to local ice flow direction, and also direction of vertical, taken from Summit Station, Greenland.}
\label{fig:GRIPCOF}
\end{figure}

The Summit Station site is among the best studied glaciological sites in the world. The original drilling and extraction of the 3027-m GRIP and 3053-m GISP-2 cores in the 1990's was followed by comprehensive conductivity and chemistry analysis \citep{taylor1993electrical}. This has been complemented by aerial and ground-based radar sounding \citep{paden2010ice}, providing both extensive layering and ice thickness information in the vicinity of the Summit Dome. Relevant to the measurements outlined below:
\begin{itemize}
    \item[--] As revealed by aerial surveys, the bedrock topography in the vicinity of Summit is observed to be highly faceted \citep{paden2010ice}.
    \item[--] Consistent with expectations for a dome, ice flow is minimal at Summit, with measured velocities of order 1~m/year in the direction of Grid West \citep{hawley2020greenland}. 
    \item[--] Ice fabric measurements indicate that the \^{c}-axis (defined as a normal to the englacial ice crystal face) distribution, projected onto the horizontal plane is approximately isotropic in azimuth (i.e., ``uniaxial'', such that the horizontal directional eigenvalues of the fabric orientation ellispoid E1 and E2 are approximately equal, but differ from the vertical eigenvalue E3), consistent with the small ice flow at the top of a dome. The \^{c}-axis alignment with vertical grows approximately linearly with depth\citep{thorsteinsson1997textures}; by a depth of 3~km, the \^{c}-axis distribution is almost entirely verticlly aligned.
\end{itemize}
The last two considerations suggest that the refractive indices for signals propagating vertically, but with polarization pointing in any arbitrary direction in the horizontal plane are approximately uniform ($n_1\approx n_2$, where $n_1$ and $n_2$ are the refractive indices for signals with polarizations pointing parallel and perpendicular, respectively, to the local ice flow direction in the horizontal plane).
This uniaxial behavior contrasts with the biaxial ice fabric at South Pole, for which the three refractive indices differ in magnitude by up to 0.2\% \citep{jordan2019modeling}. Figure \ref{fig:GRIPCOF} \citep{thorsteinsson1997textures} shows the published eigenvalues of the crystal-orientation fabric ellipsoid, illustrating the near equivalence of the E1- (parallel to ice flow) and E2- (perpendicular to ice flow, in the horizontal plane) eigenvalues and the increasing dominance of the E3- eigenvalue (\^z-direction) with depth, as compression effects increasingly rotate the \^{c}-axis towards the vertical.

\section{Experimental Configuration}
\begin{figure}
\includegraphics[width=\columnwidth]{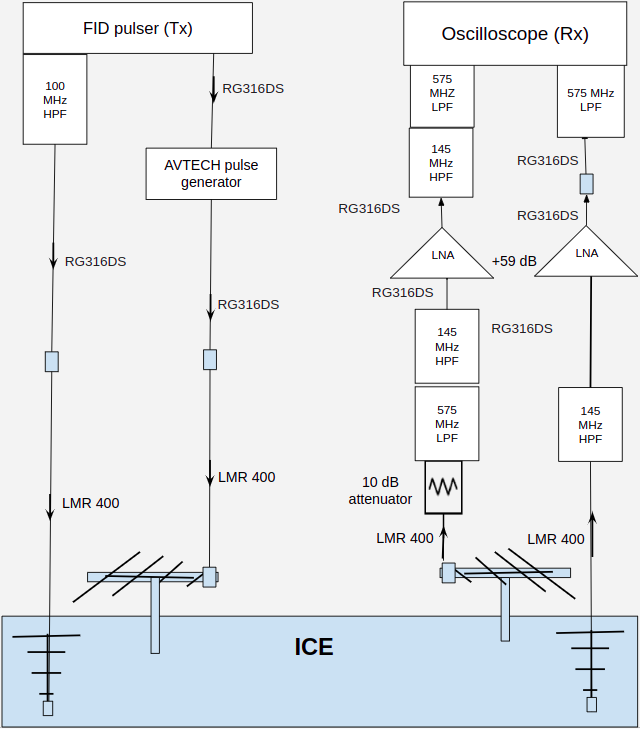}
\caption{Set up for data-taking at Summit Station, Greenland during August, 2021. Signal chain is as follows: FID, Inc. FPG6-1PNK Signal Generator at the highest amplitude setting $\to$ Mini-Circuits, Inc. NHP-100 100 MHz high-pass filter $\to$ 10m LMA-400 coaxial cable$\to$ RNO-G log-periodic dipole antenna (LPDA) $\to$ Ice Reflector $\to$ RNO-G LPDA $\to$ 10m LMA-400 coaxial cable $\to$ VHF 145+ high-pass filter $\to$ RNO-G Surface Amplifier ($\sim$59 dB gain) $\to$ NLF 575+ low-pass filter$\to$ Tektronix digital oscilloscope for data recording at 5 GSa/s.}
\label{fig:GBS.png}
\end{figure}

Data were taken using the same experimental configuration as recently reported for the in-ice radio-frequency attenuation length, calculated by normalizing to through-air propagation (Figure \ref{fig:GBS.png}, reproduced from \citep{aguilar2022situ}). Schematically, the set-up is typical of bi-static radar experiments, in which a high-gain transmitter beams signal downwards into the ice and the reflected signals are recorded in a similar, downward-facing high-gain receiver antenna, connected to a low-noise amplifier (LNA), and horizontally displaced by $\sim$244 m. To improve the signal coupling to the surface snow, both transmitter and receiver antenna were buried approximately 2 meters. The signal trigger is provided by a direct above-ice transmitter$\to$above-ice receiver antenna. In our case, we use a sharp ($\sim$1 ns risetime) pulsed output radar signal rather than the $\sim$microsecond-duration fixed frequency tones typically employed for mapping ice thickness and bedrock topography, to directly elucidate sharp internal features without the need for advanced radar signal-processing techniques such as synthetic aperture radar. 

Since the return signal strength may depend on the electric field polarization direction, four additional data sets (analyzed for the birefringence measurement) were also taken for which the transmitter and receiver antennas were placed on the surface (again, facing down) and moved to a separation $\sim$10\% of the original separation, with the trigger provided by the through-air `leakage' from the transmitter side lobe to the receiver. Each of those four data sets corresponds to azimuthal antenna polarizations co-rotated in 30 degree increments (designated as $\phi_0$, $\phi_{30}$, $\phi_{60}$, and $\phi_{90}$) from parallel to the local ice flow direction, to perpendicular to the local ice flow direction. Given the axial symmetry of the horn antennas, this angular range should bracket the full range of possible polarization orientations.

The detailed signal chain is as follows: a FID model FPG6-1PNK signal generator (the same unit described in \citep{Besson:2010ww}) set to maximum signal amplitude is connected to a high-pass filter (Mini-Circuits NHP-100) to inhibit potentially damaging reflections due to poor impedance mismatch outside of the system bandpass. 
An additional 10-m of LMR-400 coaxial cable is then connected to an RNO-G Log Periodic Dipole Antenna (LPDA, model Create-clp-5130-2n) directed downwards into the snow. At the receiver end, an identical, downward-pointing LPDA, with the receiver antenna plane oriented parallel to the transmitter (Tx) antenna plane (i.e., co-polarized transmitter and receiver orientation) feeds signal into a +59-dB custom amplifier, and, finally, to a Tektronix digital oscilloscope, for digitization and storage.

\section{Experimental Results}
\subsection{Coherence of in-ice scattering}

In addition to coherent scattering expected from  discrete conducting layers, radar signals may be reflected incoherently by individual atomic scatterers much smaller than the characteristic radar signal wavelength. Rayleigh scattering~\citep{rayleigh1881x} over the entire ice volume formally describes the case in which the signal wavelength is much larger than the size of the scatterer, leading to incoherent $1/r^4$ amplitude dependence with distance, and scattering angles increasing with frequency (and therefore often out of our signal beam). 

Detailed first-principles calculations of the volumetric Rayleigh scattering expected for airborne ice-penetrating radar~\citep{davis_moore_1993} concluded that, although surface and layered scattering dominates in the low-elevation regions of Greenland, e.g., volume scattering should dominate airborne radar surveys over the interior plateau (East Antarctica, Dome C, Vostok, and South Pole, e.g.). Calculations of radio-frequency attenuation lengths using the `slopes' of radar data echograms typically exclude consideration of volume contributions~\citep{KennyJoeMac2012,stockham2016radio}, although the validity of this assumption that volume scattering can be neglected has not yet been rigorously demonstrated.\footnote{We note that attenuation length determinations based on bedrock echoes implicitly include any processes (either due to the bulk volume or discrete layers) which scatter signal out of the main beam.}
The incoherent volume contribution is not only of interest glaciologically, but also important for experiments seeking to measure neutrinos above some radio-frequency background using radar echoes \citep{prohira2021radar}; some of that background may be due to intrinsic thermal, black-body radiation, and some may be due to incoherent volume scattering.

\subsubsection{Coherent vs.\ incoherent scattering}

\begin{figure}
\includegraphics[width=\columnwidth]{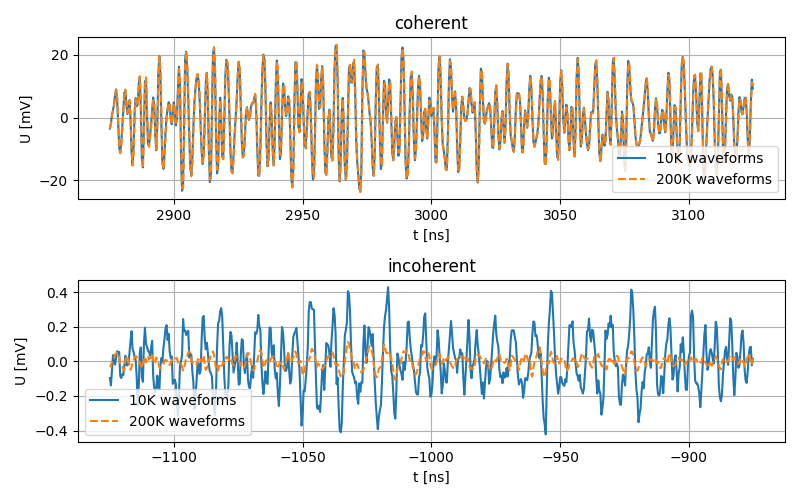}
\caption{Comparison of the averages from 10,000 waveforms ($N_{\rm files}=1$; blue) vs. 200,000 waveforms ($N_{\rm files}=20$; orange) inside a time window dominated by coherent waveform summation (top) and one dominated by incoherent noise (bottom).}
\label{fig:summed_waveforms}
\end{figure}

To numerically quantify the relative contributions of coherent (layer scattering, e.g.) vs.\ incoherent (including volume scattering) signals, 20 separate runs, each comprising an average of 10000 individual echo waveform triggers, were taken at Summit Station. For each run, a single data file, consisting of the 10K-averaged waveform, was written to the digital oscilloscope memory. Experimentally, for purely coherent/incoherent signal summation, the phase of the return signal observed at a given time $t$ (measured simply as a voltage at a given time) relative to the data-acquisition trigger time ($t_0$) is identical/random trigger-to-trigger. 
For N summed waveform files, coherent scattering amplitudes therefore scale linearly with the number of files co-added ($N_{\mathrm{files}}$) whereas incoherently summed waveform amplitudes vary as $\sqrt{N_{\mathrm{files}}}$; the latter corresponds to the expectations for summing thermal noise. Similarly,
averaging waveforms for N runs will not change the amplitude of a coherent signal, while the rms for incoherent signals will decrease by a factor $\frac{1}{\sqrt{N_{\mathrm{files}}}}$. This is demonstrated by the fits in Figure \ref{fig:summed_waveforms}, comparing the averaged waveforms for the two specified (pre-trigger `control' incoherent and immediately post-trigger `control' coherent) time intervals. 
This statistical contrast between a coherent vs. incoherent average allows us to quantify the relative contribution of coherent layer scattering to the return echoes for our data sample.

\begin{figure}
\centering
\includegraphics[width=\columnwidth]{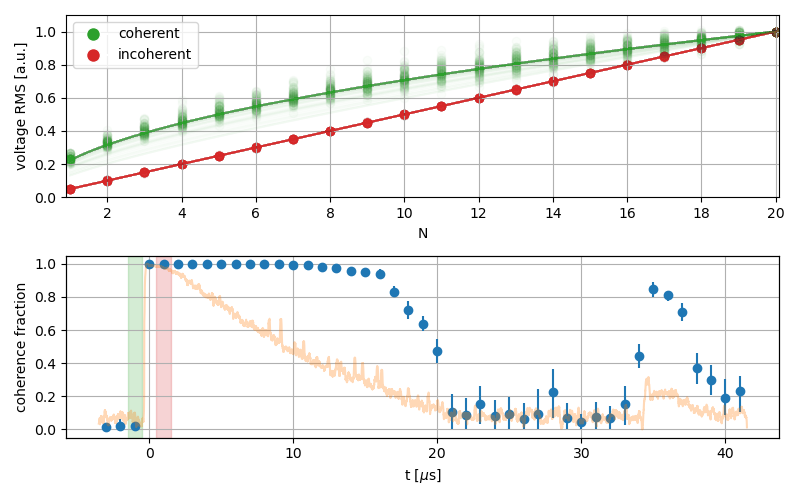}
\caption{Top: Variation in the root mean square voltage ($\sigma_V(N_{\mathrm{files}}$)) of the sum of the waveforms from N files of 10000 triggers each, with fits to a linear (red) or square root (green) dependence overlaid. Green shows the result for pure, incoherent noise, red for a time window where the amplifier was saturated and recorded identical voltages event-to-event. Scatter in the green points shows the variation obtained by shuffling the order by which runs are added to the average, to ensure no time-dependent bias in the result. Bottom: Estimated (pre-corrections; see below) relative contribution of coherent scattering to the return signal as a function of signal travel time. The green and red shaded areas mark the time window used for the plot above. For reference, the logarithm of the return power of the return signal is overlaid in orange.}
\label{fig:fCohInc}
\end{figure}

During data-taking, the digital oscilloscope recorded approximately 50 microseconds of data, including 4.5 microseconds prior to the event trigger, and approximately 10 microseconds beyond the deepest observed echo, corresponding to the bedrock at a depth of $\approx$3 km.
To apply the approach outlined above to the full waveform capture, we separate our observed echograms into four time regimes. For t $< 0~\mu$s, i.e.\ before the first signals from the transmitting antenna arrive prior to the t$\approx$0 trigger signal, data are dominated by noise from the environment, which can serve as a control region for entirely incoherent signals. Some of the radio signal broadcast from the downwards-facing transmitter antenna leaks sideways and, via a through-air route, is strong enough to saturate the nearby receiver antenna over the time interval 0 $\mu$s < t $\lesssim$ 5 $\mu$s. Over this time interval, recorded Rx voltages are dominated by the ringdown of the saturated amplifier, which are identical trigger-to-trigger, and can therefore be used as a control coherent region. This is followed by return signals from any reflectors inside the ice (either coherent or incoherent), until the noise floor is reached at $t_{\mathrm{echo}}\sim$20$\mu$s.

We measure the relative contribution of coherent scattering $f_{c}$ to the total signal by calculating the root mean square $\sigma_V (N_{\mathrm{files}})$ of the sum of the waveforms for N files, which is shown for the two control regions in Figure \ref{fig:fCohInc}. 
We now solve for the coherence fraction at an echo time $f_c(t_{\mathrm{echo}})$ by fitting a function of the form
 \begin{equation}
     f(N) = f_{c}(t_{\mathrm{echo}}) \cdot \frac{N}{20} + (1 - f_c(t_{\mathrm{echo}})) \cdot \sqrt{\frac{N}{20}}
 \end{equation}
to the $\sigma_V (N_{\mathrm{files}})$ distribution, normalized to $\sigma_V (20) = 1$, in slices of echo time $t_{\mathrm{echo}}$, to extract the relative coherence fraction $f_c$ for that echo time slice. For one of our transmitter/receiver polarization orientations, the initial result of this exercise is shown in Figure \ref{fig:fCohInc}, bottom. Our experimental data indicate completely coherent scattering for the first 15 microseconds, after which signal observed amplitudes degrade and become comparable to the $t<0$ incoherent control region, albeit wth a slight positive offset. 

\subsubsection{Cross-checks}
We have performed several cross-checks of the result presented above, as follows:

\begin{enumerate}
    \item Data-driven, vs. parameterized fitting procedure: In addition to fitting our data as described above, we can also perform a bin-by-bin $\chi^2$ fit to the $\sigma_V(N_{\mathrm{files}})$ distributions, for a given bin of echo return time. In this case, we use our `control' incoherent vs. coherent histograms (Figure \ref{fig:fCohInc}, top) as inputs, and, for an arbitrary time slice, directly fit to the linear sum of the two histograms that gives the best match to the observed $\sigma_V(N_{\mathrm{files}})$ distribution for a non-control echo time bin. This procedure is observed to give essentially identical results to the parametric fitting approach outlined above. 
    \item Dependence on number of sub-samples used for fitting. Rather than fitting our data using 20 data points (one per file), we can combine data samples pairwise and refit to 10 points of 20K events per point. As expected, this gives consistent results - the smaller number of data points is compensated by the smaller scatter point-to-point.
    \item Dependence on return echo binning. We have verified that binning our results in one microsecond echo return bins vs.\ 250 nanosecond echo return bins yields consistent results.
    \item To allow for the possibility that the low-noise amplifier gain may slightly change over time, or that a data file may have been contaminated by background triggers, we repeated the fitting procedure by shuffling (i.e., randomizing) the order by which each of the 20 files were added to the average. This resulted in no change in our final coherence plot.
    \item Check of amplitude independence of $f_C\equiv 1$. We have verified that our procedure returns complete coherence, independent of amplitude, by inputting the same 10K data file 20 times to our fitting algorithms and recovering 100\% coherence over the entire duration of the echogram, as expected. 
    \item We have extracted the fractional coherence using a phase-variation technique for which we measure the rms spread $\delta_{\mathrm{V}}$ between the measured voltages, for a given time bin, run-to-run; we use the Voltage recorded at a given time and a given trace as a proxy for the phase of the signal. As outlined previously, in the case of complete coherence, the measured phase/voltage at a given time should be identical run-to-run; in the case of incomplete coherence, the measured phase/voltage at a given time should vary randomly. To map the true $\delta_{\mathrm{Phase,true}}$ to our observed $\delta_{\mathrm{Phase,measured}}$ distribution, we use a toy Monte Carlo, which allows us to vary the relative input fractions of incoherent Gaussian noise and a coherent sinusoid, and determine a correction function $R_C^{\mathrm{true}}(\delta_{\mathrm{Phase,measured}})$. (We have verified that any arbitrary coherent function which has the same simulated voltage for each time bin in the simulated data file, other than a sinusoid, gives identical results.) We find fair consistency between our results for four different data sets, corresponding to data collected at the four different azimuthal polarization orientations.
    
\end{enumerate}

\begin{figure}
    \centering
    \includegraphics[width=\columnwidth]{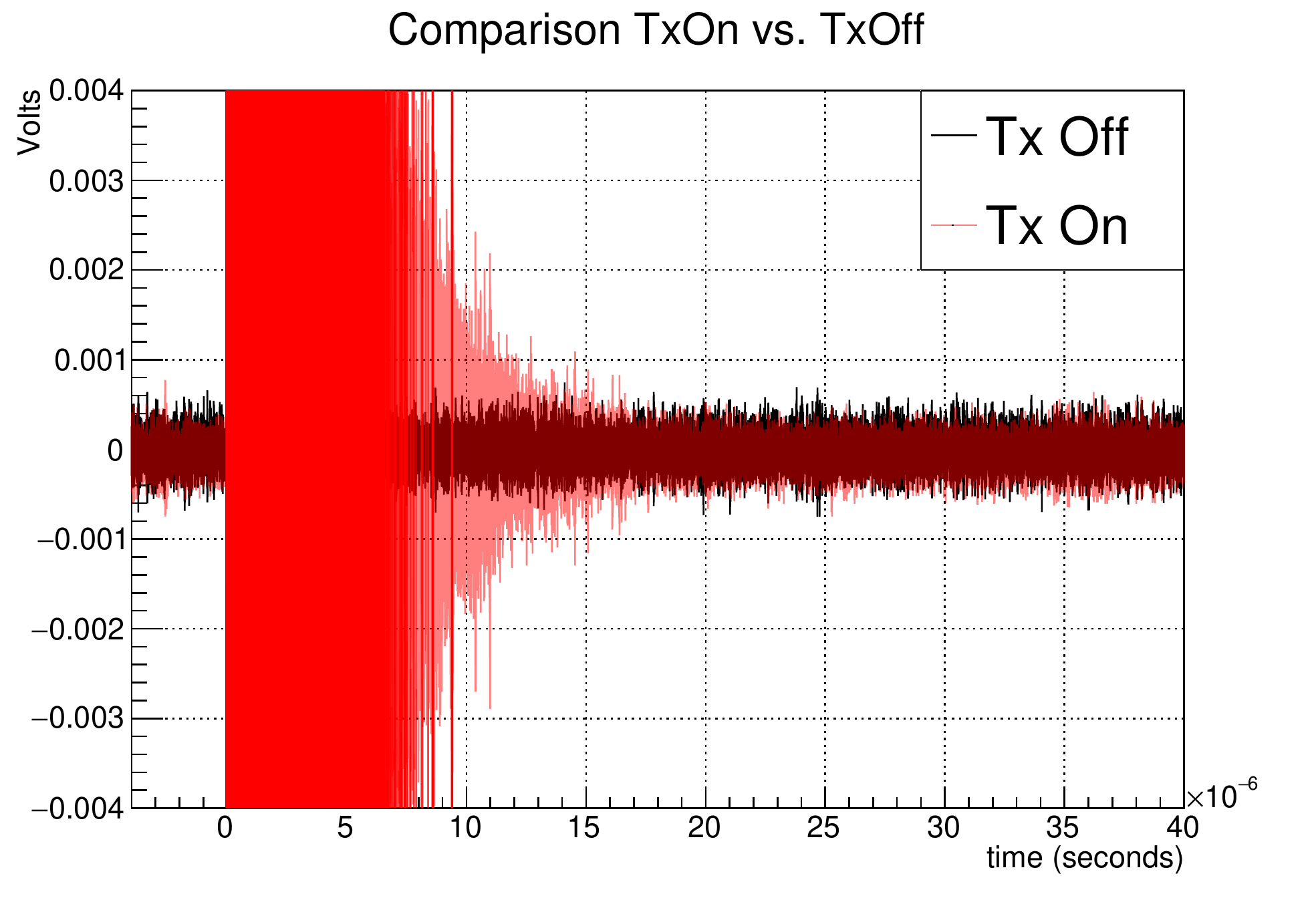}
    \caption{Comparison of traces (10K trigger averages) recorded for transmitter off (black) vs. transmitter on (red); we qualitatively note approximate consistency of the rms voltages in the 20$\to$35 microsecond time interval, as well as prior to the event trigger, at negative times.}
    \label{fig:TxOnvTxOff}
\end{figure}

In principle, the 15 microsecond time interval preceding the bedrock echo can have contributions from both volume scattering and also thermal noise. To distinguish these two, we note that: a) the expected inverse quartic dependence of volume scattering with distance would imply a depth-variable contribution varying by approximately an order of magnitude over this echo interval, inconsistent with our data, which shows a relatively flat dependence on echo time, b) A direct comparison of data taken while the transmitter was off compared to data taken while the transmitter was on shows approximately identical rms voltages over this interval, as well (Figure \ref{fig:TxOnvTxOff}). 
Performing an absolute calculation of the expected level of thermal noise, from $P_{\rm thermal}=k_BTB$ (here, $k_B$ is Boltzmann's constant, T is the ambient temperature in Kelvin, taken to be 250 K and B the bandwidth of our system, taken to be 400 MHz) also yields an expectation for a thermal noise contribution within approximately 25\% of observation.

\subsubsection{Corrections}
\begin{figure}
\centering
\includegraphics[width=\columnwidth]{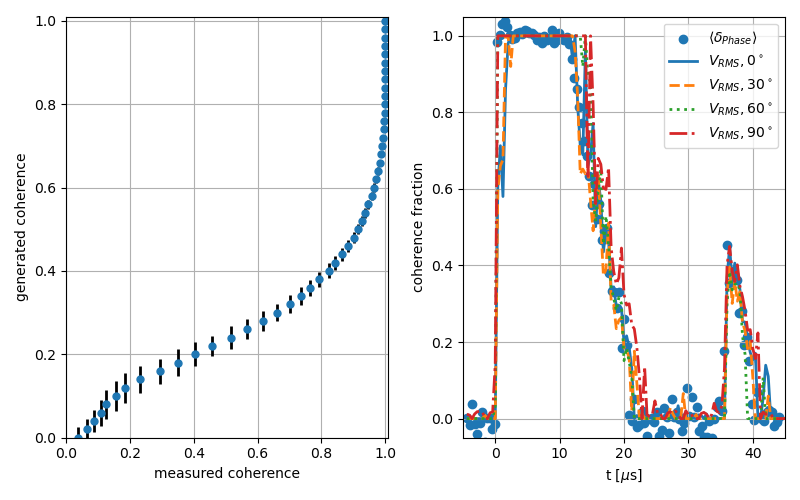}
\caption{Left: Generated coherence ($f_c$) vs. measured coherence, as obtained from Monte Carlo simulations. Right: Coherence fraction $f_c$, after applying toy Monte Carlo-based corrections, as described in text to coherence extracted using data-driven fit, without constraining relative incoherent/coherent fractions to be bounded by (0,1). Overlaid are results extracted using bin-by-bin phase coherence, from four data samples (taken with varying horizontal polarization angles referenced relative to the local ice flow direction, as indicated in the legend), as well as after applying a correction to the $f_c$ value extracted using the procedure described above, to obtain Figure \ref{fig:fCohInc}. We note some bins where the extracted coherence fraction either exceeds 1 or is less than zero. We interpret the scatter about those physical limits as indicative of the systematic errors inherent in our procedure.}
\label{fig:fCohA}
\end{figure}

Our calculation of the fractional coherence assumes a linear interpolation between the zero-coherence and the full-coherence control samples. I.e., although we have verified that we measure $f_c\equiv 1$ for a completely coherent data sample, and $f_c\equiv 0$ for a completely incoherent data sample, it is not necessarily the case that 50\% coherence will yield $f_c=0.5$. 
Using the same toy Monte Carlo as described above, we can similarly determine a correction function, that we can apply to the extracted $f_c(\sigma_V(\rm{N}_{files}))$ distribution shown in Figure \ref{fig:fCohInc}. Figure \ref{fig:fCohA}, left shows the mapping from an input $f_c$ value in our simulation to a measured value. After applying a correction based on that mapping, we obtain the results presented in Figure \ref{fig:fCohA}. Qualitatively, we obtain $f_c$ values similar to Figure \ref{fig:fCohInc}; the primary difference is that the corrected peak bedrock coherence is reduced by approximately 50\%. For echo times up to 15 microseconds, we verify nearly complete coherence in the echo returns.

Alternatively, rather than extracting the coherence using the rms of the linear sum of the coherent and incoherent voltage components in our Monte Carlo simulation, we can also consider a quadrature voltage sum as $\sigma_V(\rm{N}_{files})=\sqrt{\sigma^2_{V,coherent}+\sigma^2_{V,incoherent}}$. This leads to a linear dependence of measured $f_c$ value to input $f_c$ value in our simulation, and, ultimately, the same measured coherence fraction, as a function of time.

Our procedure indicates complete coherence ($f_c\equiv 1$, after averaging 10K triggers per file) over the echo return interval for which our measured signals significantly exceed the irreducible noise contribution, corresponding to return times less than 15 microseconds.

Three remarks are important here:
\begin{itemize}
\item $\bullet$ A higher-power transmitter would likely have greater depth reach in probing coherence and be capable of extending the coherence regime beyond 15 microseconds, or yield a larger measured bedrock coherence, e.g. The results presented herein therefore can be interpreted as a lower limit on the coherence fraction.
\item $\bullet$ Our results are specific for our data averaging, as well. We cannot exclude the possibility that volume scattering significantly contributes to the time interval over which we measure complete coherence (t$<$15 $\mu$s), since that component has already been reduced by a factor of 100 owing to the 10K averaging, prior to recording a waveform. 
\item $\bullet$ For the bedrock, we stress that our measured coherence is distinct from the intrinsic reflectivity of the bed -- in our data, for example, we observe complete coherence for shallower layers with very small intrinsic reflectivities (discussed later).
\end{itemize}
\subsection{Check of frequency-dependent attenuation length using absolute calculation.}

\begin{figure}
\centerline{\includegraphics[width=\columnwidth]{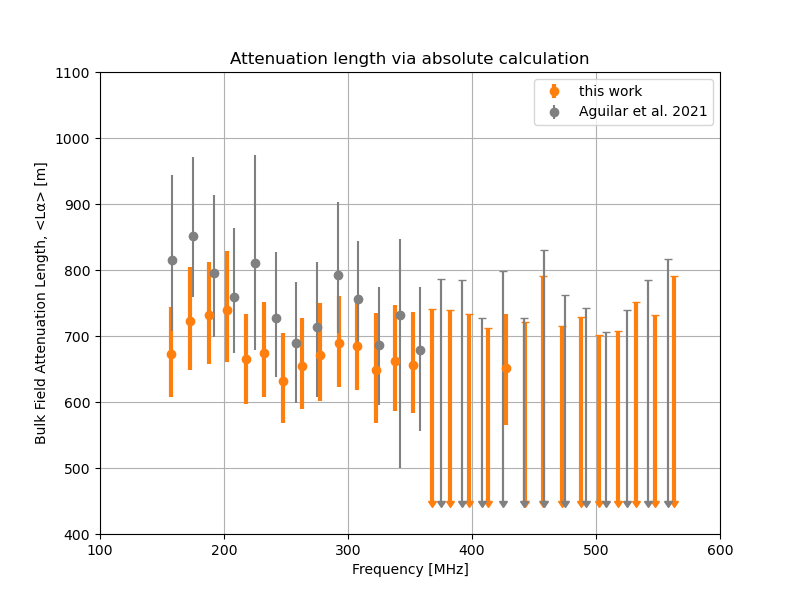}}
\caption{Calculated attenuation length, as a function of frequency, obtained by 
Friis calculation of bedrock echo power measured in a receiver Log-Periodic Dipole Antenna (LPDA) relative to the calculated signal power transmitted into the ice sheet by an identical LPDA, at Summit Station, Greenland. The prior results obtained using the in-air normalization are overlaid.}
\label{fig:bryanLatten}
\end{figure}

We now describe an `absolute' attenuation length measurement, based solely on the strength of the observed through-ice, bedrock-reflected signal, combined with our laboratory-derived values for the system transfer function. This provides a partially-independent check on the recently published measurement \citep{aguilar2022situ} derived by normalizing the observed in-ice bedrock reflection to a through-air $Tx\to Rx$ broadcast. Note that one of the primary uncertainties in that prior measurement (the bedrock signal reflectivity $R$) is inherent in this measurement, as well. 
As previously outlined \citep{aguilar2022situ}, the `absolute' extraction of the attenuation length can be derived from standard radar expressions.
The Friis Equation \citep{shaw2013radiometry} prescribes the free space wavelength-dependent received power $P_{Rx}$ given the power output from a transmitter $P_{Tx}$, over the frequency band overlapping with the receiver frequency response, for propagation over a distance $d$ through a medium with frequency-dependent average field attenuation length $\langle L_\alpha \rangle$:

\begin{equation}
P_{Rx}(\lambda) = P_{Tx}(\lambda){\cal G_{\rm Tx}}{\cal G_{\rm Rx}}{\cal G_{\rm LNA}}RF \left(\frac{\lambda}{4\pi d}\right)^2 \exp\left(-\frac{2d}{\langle L_\alpha\rangle}\right);
\end{equation}

\noindent here, $F$ is a flux-focusing factor, $R$ the bedrock reflectivity, $\lambda$ the wavelength being broadcast (since the signal is impulsive, we transform into the frequency domain and calculate transmission amplitude in frequency bins), ${\cal{G_{\rm LNA}}}$ the gain of the low noise amplifier, and ${\cal G_{\rm Tx}}$ and ${\cal G_{\rm Rx}}$ refer to the gain of the transmitter and receiver antenna, respectively.

To quantify the bedrock echo signal power, we use the same power integration window (0.5~microseconds) used in the companion analysis \citep{aguilar2022situ}. The numerical values for the quantities needed to extract the attenuation length are similarly taken from the companion paper; the primary difference is that the gain characteristics of the amplifiers and losses in cables and connectors must be explicitly inserted as multiplicative corrections to the value of $P_{Tx}$ input to the equations above, whereas in the previous measurement these systematics largely cancel.
Applying the Friis equation to the frequency-binned signal power, we obtain the attenuation length, as a function of frequency, as shown in Figure \ref{fig:bryanLatten}. 
As in the companion paper, we use a Monte Carlo method, where each parameter is varied randomly within its uncertainties and the total error estimated from the resulting distribution of attenuation lengths. We draw uncertainties on the bedrock reflectivity $R$, the signal propagation distance $d$, and the focusing factor $F$ from the previous publication. The relative uncertainty on the gain of the low noise amplifier is estimated as $\sigma_{\mathrm{LNA}}=0.1$, based on \citep{aguilar2021design}, and the gain of the antennas, which are identical to those used by the ARIANNA experiment, has an estimated relative gain uncertainty  $\sigma_\mathcal{G}=0.15$ \citep{Barwick:2016mxm}.

We obtain final values are typically within 10\% of results previously reported using the through-air normalization technique \citep{aguilar2022situ}, albeit consistently lower. We consider this level of agreement adequate to justify applying this calculational machinery to a measurement of internal layer reflectivity (as described below), which was one of the original primary calibration science goals.

\subsection{Crosscheck of attenuation measurement using envelope of return power profile}

\begin{figure}
    \centering
    \includegraphics[width=\columnwidth]{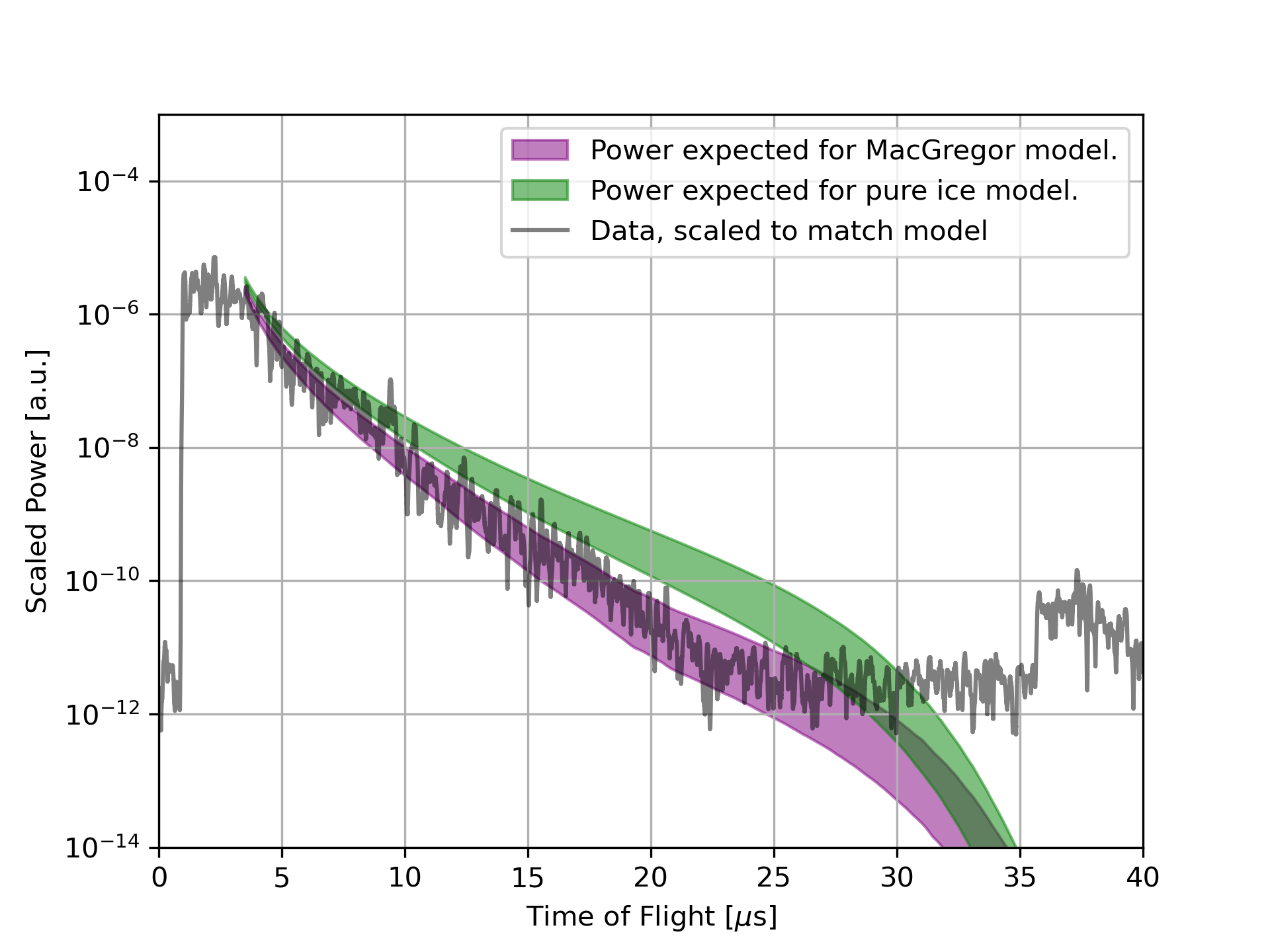}
    \caption{Measured return power of the reflected radio signal overlaid with expectations for attenuation models based on \cite{KennyJoeMac2012}, if impurities are included (purple) or ignored (green).}
    \label{fig:layer_attenuation}
\end{figure}

The measurement shown in the previous section measures the total attenuation through the entire thickness of the ice sheet; the depth dependence must be inferred based on ice core data. This is especially unfortunate, as the bulk attenuation is expected to be dominated by the deepest \SI{\sim 1}{km} of the ice sheet while the most relevant region for radio neutrino detectors is the upper \SI{\sim 1.5}{km}, where the majority of detectable neutrino interactions are expected to occur.

However, we can use the power envelope from in-ice scattering as a cross-check, which should scale with distance as $1/r^2$ for coherent scattering; any additional loss can therefore be attributed to attenuation. Having measured the total two-way attenuation to the bedrock and back, we invoke a model of relative attenuation, as a function of depth to infer the total attenuation to an arbitrary depth within the ice, as detailed in the companion measurement\citep{aguilar2022situ}. Given that conducting impurities should contribute significantly to the total attenuation, and that the impurity concentrations as a function of depth are known, we can therefore use these data to check our attenuation length calculation, expecting that a model that incorporates impurity effects should provide a better match to observation.
This approach is complementary to the ground echo measurement, as it is not affected by the deeper parts of the ice sheets and very little affected by systematic uncertainties in the instrument response.

\cite{MacGregorModel} present a model for the depth dependence of the attenuation length, which takes into account the conductivity variation with temperature of impurities in the ice. Figure \ref{fig:layer_attenuation} shows the measured return power as a function of echo time, calculated by integrating the square of the voltages over a \SI{100}{ns} sliding window. Superimposed is the expectation using the MacGregor {\it et al.} model for attenuation, as a function of depth. The measured return power has been rescaled to match the expectation at the early times, when the amplifier has just recovered from saturation. For comparison, the expected return power if only the effect of ice temperature is considered, but impurities are ignored (referred to as the pure ice model) is also shown. The pure ice model predicts a longer attenuation length (around \SI{1100}{m} at \SI{150}{MHz}) at shallower depths compared to the MacGregor model (around \SI{850}{m} at \SI{150}{MHz}), for which impurities in the ice decrease the attenuation length over the upper \SI{\sim 1500}{m} of the ice sheet. 
We observe that the MacGregor model qualitatively matches the shape of the measured return power with depth much better than the pure ice model (other models, such as (\cite{paden2010ice} give predictions similar to MacGregor {\it et al.}). These results therefore indicate internal consistency between our measured attenuation length dependence on depth, and the shape of the return echo power envelope measured in our data.

\subsection{Internal layer echo strength and calculation of internal layer reflectivity}
If the reflectivity of internal layers is sufficiently high, the amplitude of radio signals generated by ultra-high energy cosmic ray air showers impacting the snow surface, and subsequently reflected upwards to a near-surface radio receiver, may present an irreducible background to neutrino searches \citep{DeKockere:2021qka,rice-smith_2022}.

\begin{figure}
    \includegraphics[width=\columnwidth]{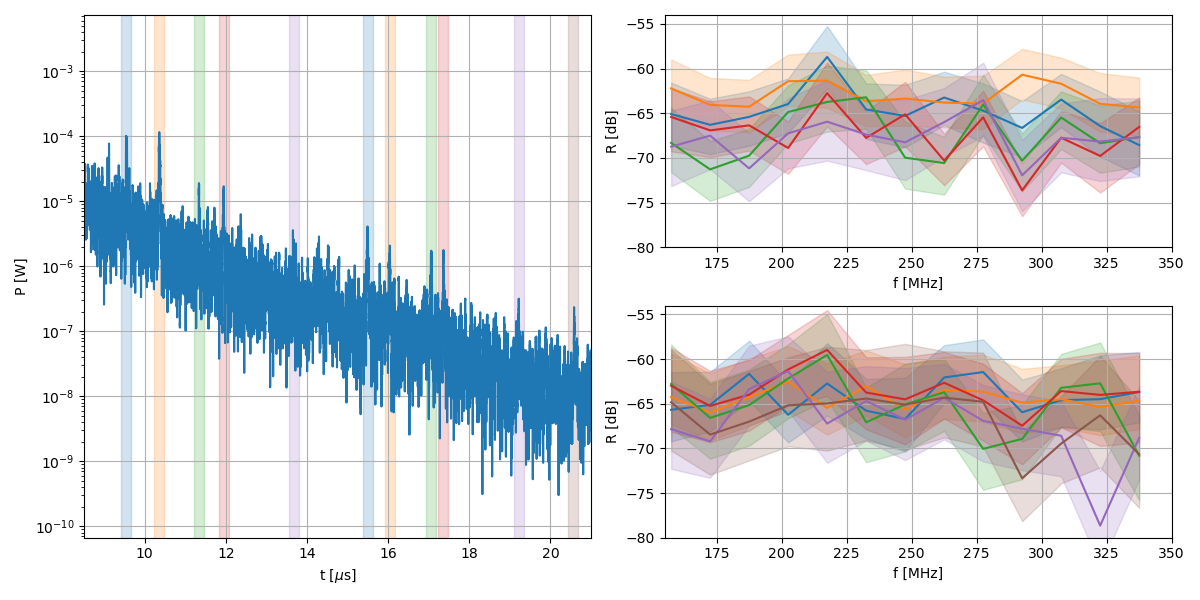}
    \caption{Measurement of the reflectivity of the most prominent reflectors in the ice. Left: Return power as a function of signal propagation time. The colored bands mark the position of the radio reflectors. Right: Reflectivity of the radio reflectors considered. For better readability, the measurements have been split between two figures, with the top plot showing the first five reflectors and the bottom plot the last six reflectors.}
    \label{fig:layer_reflectivity}
\end{figure}

We can directly apply the formalism described previously, used to numerically extract the attenuation length, to a calculation of the absolute reflectivity of the observed internal layers. In this case, rather than inputting a value for the bedrock reflectivity, as we do for the attenuation length measurement, we use the measured return strength at a given echo time to extract the reflectivity itself. We use our own measurement of attenuation length, as a function of depth, to quantify the in-ice signal absorption, down to a given layer depth\citep{aguilar2022situ}. Figure \ref{fig:layer_reflectivity}, left highlights the layers considered; the right panel shows, for each of the considered layers, the Friis-derived reflectivity as a function of frequency. Reflectivities are approximately \SI{-65}{dB}, with no clear dependence on frequency or change with depth.

Systematic errors for this absolute reflectivity measurement are assessed using a toy Monte Carlo, similar to that used for the absolute attenuation length measurement~\citep{aguilar2022situ}. The uncertainties on the antenna and amplifier gains, firn focusing, and the refractive index are identical to \citep{aguilar2022situ}; the integrated attenuation to a given layer is calculated using our own attenuation length measurement results, and also varied within corresponding errors. In our toy Monte Carlo, we ignore any possible correlated errors between the varied quantities, which yields a conservative estimate of our total uncertainties after adding all errors in quadrature. 
During the measurements to check for birefringence (see following section), we noticed that the magnitude of the return power from bulk ice reflections varied between different antenna orientations relative to the direction of ice flow by as much as \SI{40}{\%}.
Admitting the possibility that this may have been an instrumental effect, we correspondingly assume an additional \SI{40}{\%} uncertainty on the return power in our calculation of the total error.

\subsection{Polarization Dependence and Birefringence}
\label{sec:birefringence}
Unlike South Pole, for which the measured crystal orientation fabric (COF) exhibits a strong variation in the horizontal plane \citep{Voigt2017}, the measured COF at Summit Station is significantly more azimuthally isotropic, corresponding to a uniaxial, rather than biaxial, COF (Figure \ref{fig:GRIPCOF}). Consequently, we expect birefringent effects to be less evident for a vertically propagating signal at Summit Station compared to South Pole. To test this, four sets of radar echo data were compiled at $\phi_0=0^\circ$, $\phi_{30}=30^\circ$, $\phi_{60}=60^\circ$ and $\phi_{90}=90^\circ$ angles relative to the direction of the local ice flow (approximately due west, at a velocity an order of magnitude smaller than for South Pole). Biaxial birefringence can result in a $\phi$-dependent wave velocity in ice -- at South Pole, this corresponds to maximum time lag $\delta(t)\sim$10 ns per kilometer of traversed ice for echoes reflected off the bedrock \citep{Allison:2019rgg}.

To quantify birefringence, we window the summed and averaged waveforms $\pm$2 microseconds around the initial observed bedrock echo time for a) azimuthal polarization parallel to the flow axis ($\phi_0$) and b) azimuthal polarization perpendicular to the flow axis ($\phi_{90}$). We find that the value of the cross-correlation between these two windowed waveforms is maximized when we apply a time offset of +1.6 ns to the perpedicular-flow waveform. To quantify uncertainties, we perform a similarly windowed cross-correlation between all unique pairs of the twenty event files (each comprising 10K averaged events) corresponding to a single $\phi$ orientation; the distribution of time offsets peaks at 0.10 ns (consistent with zero, as expected), with a standard deviation of 3.3 ns. 
We therefore quote a birefringent time asymmetry of $1.6\pm 3.3$ ns, compared with a central value of 53 ns measured at the South Pole~\citep{Besson:2010ww}, over a 6\% shorter pathlength. Our measured value is considerably smaller than the group delay of typical radio neutrino experiments, indicating that birefringence will not be as important a consideration in reconstructing neutrino signals following primarily vertical trajectories, compared to South Pole. Additional work is needed to quantify this effect for horizontal trajectories, which admit both vertical and also horizontal polarizations.

\begin{figure}
    \centering
    \includegraphics[width=\columnwidth]{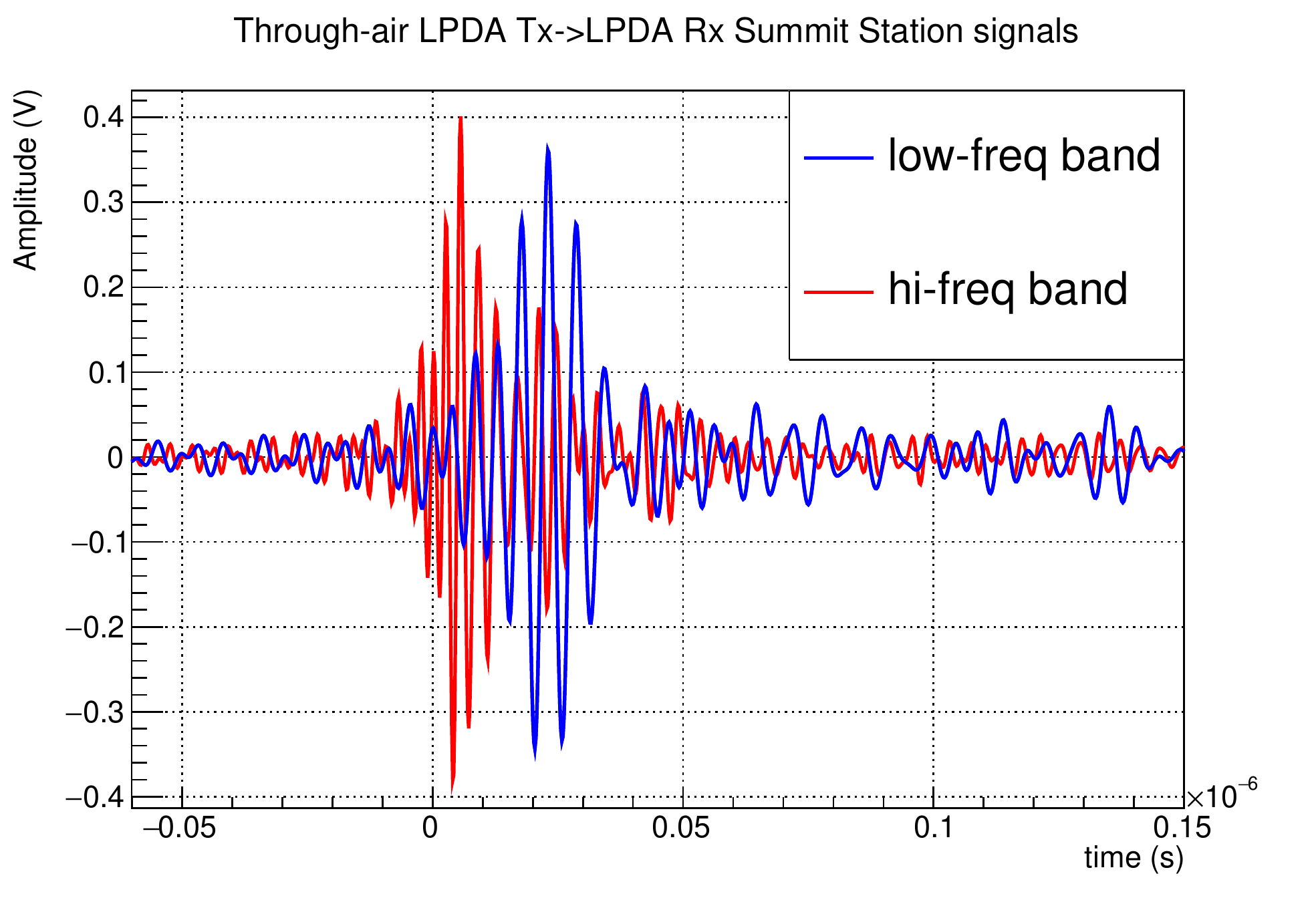}
    \caption{Frequency-banded signals, for in-air broadcasts between LPDA surface transmitter and LPDA surface receiver. The observed time shift in signal arrival times is consistent with the known, intrinsically dispersive properties of the LPDA antennas.}
    \label{fig:Dispersion_InAir}
\end{figure}

\subsection{Dispersion}
Among the ice parameters critical to the design of an in-ice radio-frequency neutrino detection experiment is the variation of refractive index over the relevant system frequency band, i.e., dispersion. Laboratory studies of pure ice indicate that the dependence of wave speed on frequency at radio-frequencies (n($\omega$)) should be insignificant over the signal propagation distances typical of experiments such as RNO-G~\citep{fujita1993measurement}. Nevertheless, there are relatively few direct {\it in situ} determinations of dispersion~\citep{Besson:2010ww}. One approach for measuring this critical parameter might be to measure the simultaneity of the bedrock echo as one sweeps in frequency, using bi-static surface radar. 

Given the $\lambda^2$ dependence of the antenna effective area, we have banded our data into two frequency bins, spanning 150-190 MHz and 190-340 MHz, respectively. Owing to the intrinsically dispersive nature of the transmitter and receiver LPDA antennas, we expect that the observed signals should show an $\sim$10-20 ns offset between the signal onset for the lower- vs.\ higher-frequency signal bands. The high frequency signal arrival for the through-air path is, indeed, observed to precede the low-frequency signal arrival by 22.5$\pm$4.2 ns (Figure \ref{fig:Dispersion_InAir}).

Cross-correlating the observed high-frequency bedrock echo with the observed low-frequency bedrock echo yields a time offset of 21.4$\pm$4.0 ns, statistically consistent with the offset observed for the through-air path. Subtracting that offset, we obtain a time difference between the two bands of 1.1$\pm$5.6 ns (adding the statistical uncertainties on the bed and through-air offsets in quadrature). Normalized to the total bedrock echo time of 35.5 microseconds implies a dispersive slope of ($3.1\pm15.8)\times 10^{-5}$/100 MHz. This value is statistically consistent with the value obtained at the South Pole~\citep{Besson:2010ww}, albeit with a larger attendant error, in part owing to the more dispersive LPDA antennas used as transmitter/receiver, and in part owing to the more restricted frequency band over which these measurements were made.

\section{Conclusions}
We summarize our results as follows:
\begin{enumerate}
    \item Using the Friis equation, we determine the radio-frequency ice attenuation length at Summit Station and reproduce (to $\approx$10\%) the values calculated by normalizing the in-ice reflection to a through-air broadcast \citep{aguilar2022situ}.
    \item Using the same approach as for the bedrock reflector, we similarly measure the reflectivity of internal layers at Summit Station. We obtain values in the range $-60\ \mathrm{dB}$ to $-70$ dB, roughly consistent with values obtained for a similar analysis at South Pole \citep{Besson:2021wmj}.
    \item Based on the statistics of our observed signals, and the dependence of the measured rms-voltage with sample size, we find no clear evidence for volume scattering within the ice sheet, after averaging $N_{\rm trigger}$=10K triggers per recorded waveforms. However, we cannot exclude the possibility that volume scattering may contribute more significantly as $N_{\mathrm{trigger}}$ approaches 1. 
    \item Consistent with expectations based on the known crystal fabric at Summit Station, Greenland, no significant birefringent effects are observed in our radar sounding data for vertically propagating signals.
    \item We observe no evidence for dispersion of radio-frequency signals propagating in ice, over the frequency interval 150-340 MHz.
\end{enumerate}

\noindent Overall, from a radio-glaciological perspective, Summit Station is an extremely propitious site for in-ice neutrino detection. During the summer of 2022, RNO-G more than doubled its sensitive neutrino volume. Given the importance of ice properties to neutrino detection, future campaigns anticipate an expansion of the calibration measurements cited herein, including measurements of birefringence along horizontal trajectories, more precise parameterizations of the refractive index dependence on depth, measurement of cross-polarized reflection strengths, as well as internal layer reflections in the upper 200 meters of the ice sheet. 

\subsection{Acknowledgements}

We would like to thank the staff at Summit Station and Polar Field Services for their unflagging and superb logistical support.
We thank our home institutions and funding agencies for supporting the RNO-G work; in particular the Belgian Funds for Scientific Research (FRS-FNRS and FWO) and the FWO programme for International Research Infrastructure (IRI), the National Science Foundation through the NSF Awards 2118315 and 2112352 and the IceCube EPSCoR Initiative (Award ID 2019597), the German research foundation (DFG, Grant NE 2031/2-1), the Helmholtz Association (Initiative and Networking Fund, W2/W3 Program), the University of Chicago Research Computing Center, and the European Research Council under the European Unions Horizon 2020 research and innovation programme (grant agreement No 805486).

\bibliography{igsrefs}
\bibliographystyle{igs}

\end{document}